\documentclass[reprint,pre,aps,superscriptaddress,showkeys,showpacs]{revtex4-1}
\usepackage{graphicx,times,CJK}
\usepackage{amssymb}
\begin{document}
\begin{CJK*}{UTF8}{}
\title{Instability development of a viscous liquid drop impacting a smooth substrate}
\author{Lei Xu (\CJKfamily{bsmi}徐磊)}
\affiliation{Department of Physics, The Chinese University of Hong
Kong, Hong Kong, P. R. China} 
\begin{abstract}
We study the instability development during a viscous liquid drop
impacting a smooth substrate, using high speed photography. The
onset time of the instability highly depends on the surrounding air
pressure and the liquid viscosity: it decreases with air pressure
with the power of minus two, and increases linearly with the liquid
viscosity. From the real-time dynamics measurements, we construct a
model which compares the destabilizing stress from air with the
stabilizing stress from liquid viscosity. Under this model, our
experimental results indicate that at the instability onset time,
the two stresses balance each other. This model also illustrates the
different mechanisms for the inviscid and viscous regimes previously
observed: the inviscid regime is stabilized by the surface tension
and the viscous regime is stabilized by the liquid viscosity.
\end{abstract}
\maketitle
\end{CJK*}
The phenomenon of a liquid drop hitting a solid surface is
ubiquitous: it occurs whenever the very first rain drop reaches the
ground or when we spill coffee onto the floor. Liquid-solid impact
has been extensively studied due to its broad applications in many
industrial processes, such as ink-jet printing, surface coating,
combustion of liquid fuel, plasma spraying, and pesticide
application\cite{rev}. It may seem obvious that the impact outcomes
should be determined by either the liquid or the solid
properties\cite{Mundo, Rioboo, Quere, Chandra, Deegan, Detlef},
however, recent studies surprisingly revealed the crucial role of
the surrounding atmosphere: reducing air pressure can completely
suppress the liquid drop splashing on a smooth substrate\cite{Xu1,
Xu3}, and the compressibility of the surrounding air is demonstrated
to be important\cite{Brenner1, Brenner2}. This unexpected discovery
brings a completely new effect, the air effect, into the impact
problem. To fully understand this new effect, therefore, it is
essential to clarify the interactions between air and the
fundamental liquid properties, such as surface tension and
viscosity. Previous study has shown that the competition between the
air effect and the liquid surface tension determines the impact
outcomes of inviscid liquid drops\cite{Xu1}. However, there has been
very limited study on the interaction between air and the liquid
viscosity, although the liquid viscosity itself has been broadly
tested\cite{Mundo, Rioboo, Wendy} and the entrapment of air bubbles
in viscous drops was illustrated recently\cite{Siggi2, Siggi3}. As a
result, the relationship between surrounding air and the liquid
viscosity is still missing, and the understanding on the
liquid-solid impacts, especially the newly discovered air effect,
remains incomplete.

In this paper, we systematically study the interaction between air
and the liquid viscosity by varying both the surrounding air
pressure and the liquid viscosity, for the impacts of viscous liquid
drops on a smooth substrate. With high speed photography, we find
that the instability produced by an impact highly depends on the air
pressure and the liquid viscosity: the onset time of the instability
decreases with air pressure with the power law of minus two, while
it increases linearly with the liquid viscosity. From the real-time
liquid motion measurements, we construct a simple model that
compares the destabilizing stress from air with the stabilizing
stress from the liquid viscous stress. The experimental results
support the picture that the two stresses balance each other at the
instability onset time. This model also predicts the existence of a
threshold viscosity, above which the system is stabilized by the
liquid viscosity, and below which it is stabilized by the surface
tension. This prediction quantitatively agrees with the previous
experiment\cite{Xu3}.

We perform all the experiments inside a transparent vacuum chamber
whose pressure can be continuously varied from 1kPa to
102kPa(atmospheric pressure). We also independently vary the liquid
viscosity by using silicone oils of very close densities
($0.92\sim0.94$ g cm$^{-3}$) and surface tensions ($19.7\sim20.5$ mN
m$^{-1}$) but different dynamic viscosities ($4.65\sim13.2$ mPa s).
We note that all our liquids wet the substrate completely thus the
wetting conditions are kept the same for all the impacts. To make
sure that identical impact conditions are achieved each time, we
release reproducible liquid drops of diameter $d=3.1\pm0.1$mm from a
fixed height, and all the liquid drops impact a smooth and dry glass
substrate at the velocity $V_0=4.03\pm0.05$ m s$^{-1}$. The impacts
are subsequently recorded by a high speed camera at the frame rate
of 47,000 frames per second.

We probe the air-liquid interaction by inspecting the instability
development during the impact: under high-speed photography, the
impact produces a thin liquid film expanding radially along the
substrate. This liquid film is stable initially, however, a small
rim shows up around the edge at a certain moment, and subsequently
develops into larger and larger undulations(See Fig.1 left column).
We believe the appearance of the rim indicates the transition from a
stable system into an unstable one, and define the moment of the rim
appearance as the instability onset time, $t_{on}$. For example, an
instant very close to $t_{on}$ is shown in the third image of Fig.1
left column. This instability onset time, $t_{on}$, measures how
fast the system goes unstable: the smaller it is, the faster the
system becomes unstable. Interestingly, $t_{on}$ has a strong
dependence on the surrounding air pressure, $P$.  The two columns in
Fig.1 show two almost identical impacts, with only different $P$: At
$P=40$kPa(left column), instabilities show up in the third image;
while at a higher pressure, $P=63$kPa(right column), they appear at
a much earlier time in the second image.

\begin{figure}
\begin{center}
\includegraphics[width=3.2in]{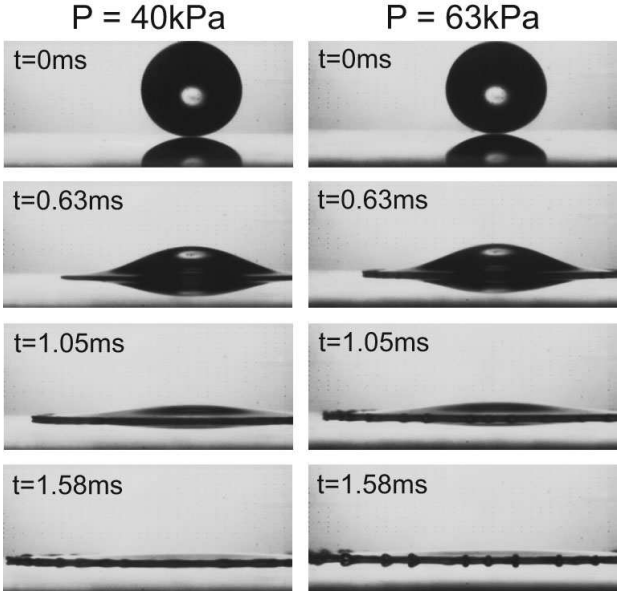}
\caption{Instability development under different pressures. The
liquid drop has diameter $d=3.1\pm0.1$mm, dynamic viscosity $\mu =
6.71\pm0.02$ mPa s and impact velocity $V_0=4.03\pm0.05$ m s$^{-1}$.
The left column shows an impact under the air pressure $P=40$kPa.
The impact is initially stable, but instability shows up from the
third image. The right column shows an identical impact under higher
pressure, $P=63$ kPa. The instability appears at a much earlier time
from the second image.}
\end{center}
\end{figure}

By performing similar experiments under different air pressures, we
systematically measure the instability onset time, $t_{on}$, with
respect to the pressure, $P$. We find that $t_{on}$ decreases
monotonically with $P$, as shown in Fig.2. Intuitively, this implies
that more air leads to earlier instability appearance, thus air acts
to destabilize the system, consistent with previous
findings\cite{Xu1}. To test the interaction between air pressure and
liquid viscosity, we perform the same $t_{on}$ vs. $P$ measurements
with silicone oils of very similar mass density and surface tension,
but different dynamic viscosities, as plotted by the different
symbols in Fig.2. From bottom to top, the four curves correspond to
increasing dynamic viscosities: $\mu$ = 4.65($\bullet$),
6.7($\circ$), 9.3($\blacktriangle$), and 13.2($\times$) mPa s.
Intriguingly, all the data can be excellently fitted by a simple
functional form: $t_{on} = A/P^2 + t_0$, with $A$ and $t_0$ the
fitting parameters. $t_0$ has typical values between 0.03 to 0.09ms,
much smaller than most $t_{on}$ values. However, it is still larger
than our time resolution(0.02ms) and can not be explained as
measurement errors. One possibility is that the system actually
becomes unstable slightly earlier than the measured $t_{on}$, but
the instability features at that moment are too tiny to visualize.
The pre-factor, $A$, increases with the viscosity, $\mu$, as
illustrated by the higher locations for larger viscosity liquids.
This result can be intuitively understood: the larger the viscosity,
the more stable the system is, and the later the instability shows
up. Limited by experimental conditions, each data set only has the
dynamic range of about one decade in time and pressure, but it is
nevertheless impressive that one simple functional form fits all the
curves nicely.

\begin{figure}
\begin{center}
\includegraphics[width=3.2in]{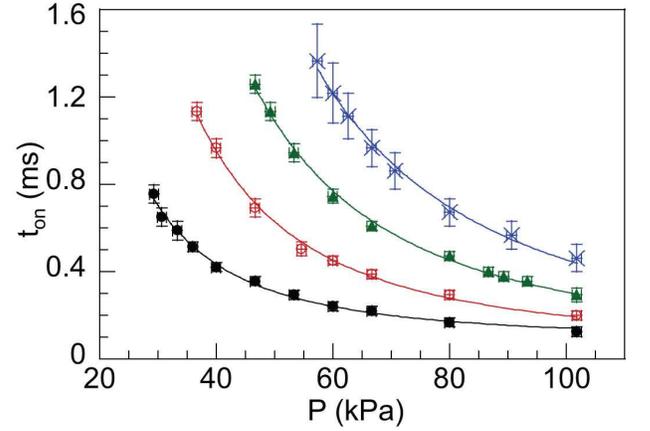}
\caption{(Color online) The instability onset time, $t_{on}$, v.s.
$P$ for liquids of different viscosities. From bottom to top, the
four curves correspond to increasing viscosities: $\mu$ =
4.65($\bullet$), 6.7($\circ$), 9.3($\blacktriangle$) and
13.2($\times$) mPa s. All the curves can be fitted by a universal
functional form: $t_{on}$ = $A/P^2 + t_0$. $t_0$ ranges from
$0.03$ms to $0.09$ms, much smaller than most $t_{on}$ values. The
pre-factor $A$ increases with $\mu$, as demonstrated by the higher
locations of the liquids with larger $\mu$. Limited by the
experimental condition, each data set has only about one decade in x
and y directions. But it is nonetheless impressive that one simple
functional form can fit all the data sets well.}
\end{center}
\end{figure}

Together these data demonstrate that the instability development
depends on both $P$ and $\mu$, but they play opposite roles: $P$
acts to destabilize the interface since higher $P$ leads to faster
growth of the instability; while $\mu$ favors stabilizing the
interface as higher $\mu$ slows down the instability growth. To
quantitatively understand the effects of $P$ and $\mu$, we inspect
their corresponding stresses: at the edge of the expanding liquid
film, air pressure applies a destabilizing stress, $\Sigma_G\sim
\rho_G\cdot C_G\cdot V_e$ \cite{Xu1}; and the liquid viscosity
produces a stabilizing stress, $\Sigma_{\mu} \sim \mu V_e /d$. Here
$\rho_G$ is the density of the surrounding gas, $C_G$ is the speed
of the sound in the gas, $V_e$ is the liquid disc expanding
velocity, and $d$ is the liquid film thickness measured at the edge.
$C_G$ enters the problem because previous experiments\cite{Xu1,Xu3}
and simulations\cite{Brenner1, Brenner2} suggest that the
compressibility of the surrounding air is important.

Since $V_e$ and $d$ vary with time, so do $\Sigma_G$ and
$\Sigma_\mu$. Therefore a careful examination on their time
dependence could provide valuable insight for the instability
development. We can directly measure $r(t)$ and $d(t)$ from
high-speed photography, as illustrated in Fig.3 upper inset. $V_e$
can be obtained by taking the time derivative of $r(t)$. Our
measurements show that $r(t)\propto \sqrt{t}$, consistent with
previous studies, thus $V_e = dr/dt\propto 1/\sqrt t$. This time
dependence keeps valid for most of the expanding period, within
which all our measurements are performed. Moreover, we can directly
measure the thickness of the liquid film, $d$, with respect to $t$,
as plotted in the main panel of Fig.3. Because the small values of
$d$ approach the single pixel level of our camera, the data are
quite discrete; nonetheless they are consistent with the fit:
$d\sim\sqrt{\nu t}$, with $\nu=\mu/\rho_L$ being the liquid
kinematic viscosity. This shows that $d$ is determined by the
boundary layer thickness, $\sqrt{\nu t}$.

\begin{figure}
\begin{center}
\includegraphics[width=3.2in]{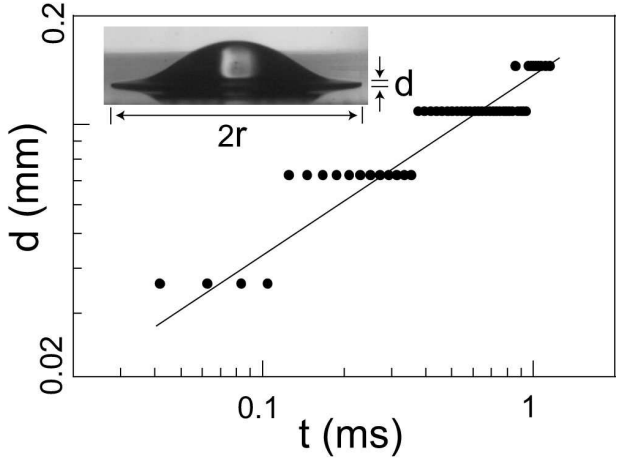}
\caption{Direct measurement of the thickness $d$ vs. time $t$. The
impact is by a liquid drop of $\mu$=4.65mPa s and
V$_0$=4.03$\pm$0.01m/s. The inset shows a typical snapshot from
which $d$ is measured: $d$ is the liquid film thickness measured at
the edge. Main panel shows the measured $d(t)$. Because $d$ is quite
small, the four discrete values correspond to one, two, three and
four pixels of our camera. The fit is: $d=1.9\sqrt{\nu t}$,
indicating that $d$ is determined by the boundary layer thickness:
$\sqrt{\nu t}$.}
\end{center}
\end{figure}

From the real-time dynamics, we derive the time dependence of the
stresses: The destabilizing stress, $\Sigma_G\sim\rho_G\cdot
C_G\cdot V_e\propto1/\sqrt t$, decreases with $t$ with the power of
$-\frac{1}{2}$; while the stabilizing stress, $\Sigma_\mu\sim\mu
V_e/d\propto 1/t$, depends on $t$ with the power of $-1$. Clearly,
when $t$ is small, $\Sigma_\mu \gg \Sigma_G$, and the stabilizing
stress dominates the destabilizing stress. This implies that the
system should be stable initially, as we have observed. As $t$
increases, however, $\Sigma_\mu$ decreases much faster than
$\Sigma_G$ and a crossover should occur at a certain time. After
this crossover time, $\Sigma_G$ becomes dominant and the system will
go unstable. The experiments are consistent with this picture: all
the impacts are indeed stable initially and become unstable after
the instability onset time, $t_{on}$. Therefore $t_{on}$ naturally
corresponds to the crossover time at which the two stresses balance
each other:

\begin{equation}
\rho_G \cdot C_G \cdot V_e  \sim \mu \frac{V_e}{d} \mid _{t=t_{on}}
\end{equation}

\noindent Plugging in the relations: $\rho_G\propto P$ and $d
\propto \sqrt{\mu t}$, with $V_e$ canceling each other on both sides
and $C_G$ being a constant independent of $P$, we reach the
expression:

\begin{equation}
t_{on} \propto \frac{\mu}{P^2}
\end{equation}

\noindent This expression successfully explains the two main
features observed in Fig.2.: (1) $t_{on}-t_0 \propto 1/P^2$ and (2)
the pre-factor of this dependence, $A$, increases with $\mu$.
Moreover, Eq.~2 further predicts that $A$ should increase linearly
with $\mu$. To test this prediction, we find $A$ for each viscosity
in Fig.2 from the best fit(the solid curves in Fig.2), and plot $A$
as the function of $\mu$ in Fig.4. Indeed, a very nice linear
dependence is observed but the line does not go through the origin;
instead, it intercepts the x-axis at the finite viscosity value,
$\mu_0=3.4$mPa s.

\begin{figure}
\begin{center}
\includegraphics[width=3.2in]{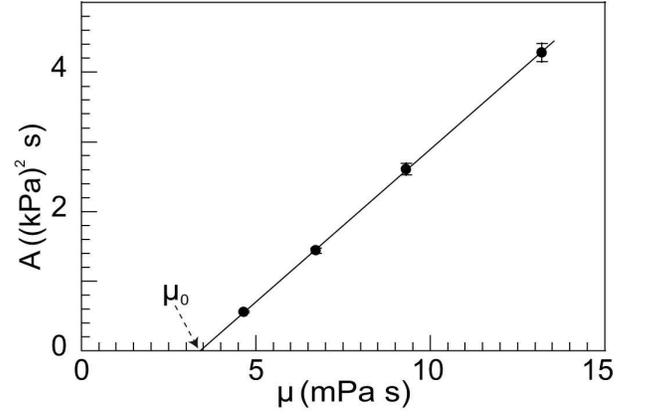}
\caption{The pre-factor, $A$, vs. liquid viscosity, $\mu$, for the
curves shown in Fig.2. The pre-factors are obtained from the best
fits in Fig. 2. $A$ varies linearly with $\mu$ and intercepts the
x-axis at $\mu_0 = 3.4$mPa S. $\mu_0$ agrees with the threshold
viscosity separating the inviscid and viscous regimes observed in
previous experiment\cite{Xu3}.}
\end{center}
\end{figure}

What is the physical meaning of $\mu_0$? To answer this question, we
need to understand the impacts by the inviscid liquids with $\mu <
\mu_0$. Previous study showed that for an inviscid liquid drop
impacting on a smooth surface, the destabilizing stress is the same
as the current viscous case, $\Sigma_G\sim\rho_GC_GV_e$ \cite{Xu1}.
However, the stabilizing stress, $\Sigma_L$, is quite different.
$\Sigma_L$ comes from the liquid surface tension, and is typically
estimated as the surface tension coefficient, $\sigma$, divided by
the liquid film thickness, $d$: $\Sigma_L\sim\sigma/d$ \cite{Xu1}.
Therefore, we propose that the complete stabilizing effect for an
impact should include both the surface tension component,
$\Sigma_L$, and the viscosity component, $\Sigma_\mu$. When the
viscosity is small, $\Sigma_L$ dominates $\Sigma_\mu$, and we get
typical inviscid behavior\cite{footnote2}. However, when $\mu$
exceeds a certain threshold value, the viscous stress $\Sigma_\mu$
will become the major stabilizing factor, and we get the currently
observed viscous behavior. Therefore $\mu_0$ naturally corresponds
to this threshold viscosity which determines whether the inviscid or
the viscous model should be used. We note that $\mu_0$ should depend
on detailed impact conditions such as the impact velocity, surface
tension and wetting conditions. Previous experiments with similar
impact conditions already confirmed that two impact regimes exist
when $\mu$ is varied, and the transition from the inviscid regime to
the viscous regime is close to $\mu_0$(see ref. \cite{Xu3} Fig.5).
This provides a strong experimental evidence for the physical
meaning of $\mu_0$. Moreover, our picture not only explains the
meaning of $\mu_0$, it also demonstrates the main difference between
the two impact regimes: the inviscid regime is stabilized by the
surface tension and the viscous regime is stabilized by the liquid
viscosity.

\begin{figure}
\begin{center}
\includegraphics[width=3.2in]{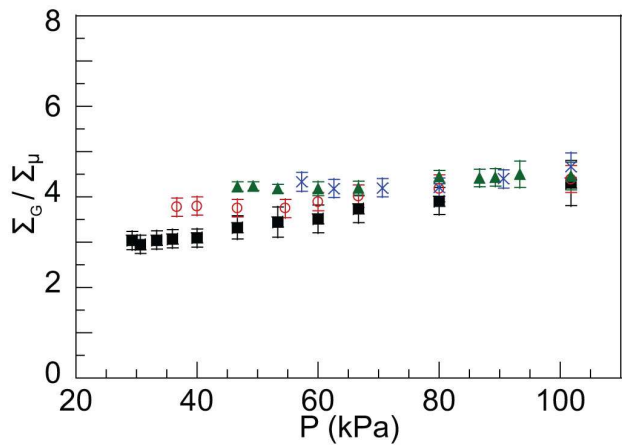}
\caption{(Color online) The ratio between the destabilizing and the
stabilizing stresses, $\Sigma_G/\Sigma_\mu$, measured at $t=t_{on}$,
for various pressures and viscosities. All experiments are done
under almost identical impact conditions, with only the pressure
being varied. Different symbols represent liquids of different
viscosities: $\mu$ = 4.65($\bullet$), 6.7($\circ$),
9.3($\blacktriangle$), and 13.2($\times$) mPa s. The ratio,
$\Sigma_G/\Sigma_\mu\sim\rho_GC_Gd/\mu$, is computed from direct
measurements: $\rho_G$ is calculated from $P$, and $d$ is from the
best fit to the high-speed images at the time $t_{on}$. Without any
fitting parameter, all the ratios are within the narrow range
between 3 and 4, confirming that $\Sigma_G$ and $\Sigma_\mu$ are
comparable at the time $t_{on}$.}
\end{center}
\end{figure}

We propose that in the viscous regime, the stabilizing stress is
mainly from the viscous stress, $\Sigma_\mu\sim\mu V_e / d$, and
construct a model which compares $\Sigma_\mu$ with the destabilizing
stress, $\Sigma_G\sim\rho_G C_G V_e$. By assuming that $\Sigma_G$
and $\Sigma_\mu$ balance each other at the instability onset time,
$t_{on}$ (Eq.1), we successfully explain the dependence of $t_{on}$
on $P$ and $\mu$: $t_{on}-t_0 = A/P^2$ and $A \propto \mu-\mu_0$,
with $\mu_0$ the threshold viscosity separating the inviscid and the
viscous regimes. However, the most critical criterion, whether
$\Sigma_G$ and $\Sigma_\mu$ are indeed comparable at $t_{on}$,
remains to be verified. To test it quantitatively, we measure the
ratio between the two stresses,
$\Sigma_G/\Sigma_\mu\sim\rho_GC_Gd/\mu$, at the moment $t_{on}$.
This ratio is tested for various pressures and viscosities, as
plotted in Fig.5. All experiments are done under almost identical
impact conditions, with only the pressure being varied. Different
symbols represent liquids of different viscosities: $\mu$ =
4.65($\bullet$), 6.7($\circ$), 9.3($\blacktriangle$), and
13.2($\times$) mPa s. For each impact, we obtain $d$ value at
$t_{on}$ from the high-speed photography
measurements\cite{footnote3}. The air density $\rho_G$ is directly
computed from the pressure $P$. The speed of sound in air at room
temperature (20$^o$C), $C_G=343$m s$^{-1}$, is a constant
independent of $P$. Plugging in all the values, we obtain the ratio,
$\Sigma_G/\Sigma_\mu$, as plotted in Fig.5. Without any fitting
parameter, most data points collapse to the narrow range between 3
and 4. These values prove that $\Sigma_G$ and $\Sigma_\mu$ are
indeed comparable at the time $t_{on}$, as our model predicts.

We study the interaction between the air pressure and the liquid
viscosity for the impact of a liquid drop on a smooth substrate. For
viscous liquids, the impact is stabilized by the viscous stress,
$\Sigma_\mu\sim\mu V_e/d$, whose competition with the destabilizing
stress determines when the system becomes unstable. By contrast, for
inviscid liquids, the stabilizing stress comes from the surface
tension, $\Sigma_L\sim\sigma/d$. Interestingly, by inspecting the
two different stabilizing stresses, we find that the liquid
viscosity plays opposite roles in them. For $\Sigma_L$ in the
inviscid regime, we have $\Sigma_L\sim\sigma/d\sim\sigma/\sqrt{\nu
t}\propto 1/\sqrt\mu$. Here larger $\mu$ leads to larger $d$ and
smaller $\Sigma_L$, thus more viscous liquids are less stable.
However, in the viscous regime, we have $\Sigma_\mu\sim\mu
V_e/d\propto\sqrt\mu$ \cite{footnote4}. Now increasing $\mu$ will
increase $\Sigma_\mu$ and make the system more stable. This
non-monotonic behavior was already observed by previous
experiments(see ref. \cite{Xu3} Fig.5) and now can be fully
understood. In summary, our study shows that the interplay between
air and liquid viscosity is crucial in determining the outcomes of
liquid-solid impacts. The viscosity plays different roles in
different regimes, and the simple intuition that a more viscous
liquid is more stable during an impact is not always valid.

We gratefully acknowledge helpful discussions with Sidney Nagel,
Wendy Zhang, Michelle Driscoll, Alexis Berges and Emily Ching. This
project is supported by RGC Research Grant Direct Allocation
(Project Code: 2060395), MRSEC DMR-0213745 and NSF DMR-0352777.


\begin{thebibliography}{99}

\bibitem{rev} A. L. Yarin, Annu. Rev. Fluid Mech., \textbf{38}, 159
(2006).

\bibitem{Mundo} C. Mundo, M. Sommerfeld and C. Tropea, Int. J. Multiphase
Flow, \textbf{21}, 151 (1995).

\bibitem{Rioboo} R. Rioboo, M. Marengo and C. Tropea, Atomization and
Sprays, \textbf{11}, 155 (2001).

\bibitem{Quere} D. Richard, C. Clanet and  D. Qu\'er\'e,
 Nature {\bf 417}, 811 (2002).

\bibitem{Chandra} N. Z. Mehdizadeh, S. Chandra and J. Mostaghimi, J. Fluid Mech., \textbf{510}, 353 (2004)

\bibitem{Detlef} Tsai \emph{et al}, Langmuir, \textbf{25}, 12293
(2009).

\bibitem{Deegan} R.D. Deegan, P. Brunet, and J. Eggers, Nonlinearity, \textbf{21}, C1-C11 (2008).

\bibitem{Xu1} L. Xu, W. W. Zhang, S. R. Nagel, Phys. Rev. Lett.,
\textbf{94}, 184505 (2005).

\bibitem{Xu3} L. Xu, Phys. Rev. E, \textbf{75}, 056316 (2007).

\bibitem{Brenner1} S. Mandre, M. Mani, M. P. Brenner, Phys. Rev.
Lett., \textbf{102}, 134502 (2009).

\bibitem{Brenner2} M. Mani, S. Mandre, M. P. Brenner, J. Fluid Mech., \textbf{647}, 163 (2010).

\bibitem{Wendy} R. D. Schroll, C. Josserand, S. Zaleski, W. W.
Zhang, Phys. Rev. Lett., \textbf{104}, 034504 (2010).

\bibitem{Siggi2} S. T. Thoroddsen \emph{et al}, J. Fluid Mech., \textbf{545}, 203 (2005).

\bibitem{Siggi3} S. T. Thoroddsen, K. Takehara, and T. G. Etoh, Phys. Fluids, \textbf{22}, 051701
(2010).

\bibitem{footnote2} We note that $\Sigma_\mu\propto1/t$ while $\Sigma_L\propto1/d\propto1/\sqrt
t$. Thus $\Sigma_\mu$ always dominates for small $t$. However,
$\Sigma_\mu$ decreases more rapidly with $t$ and can drop below
$\Sigma_L$ soon after the impact, for the case of small $\mu$.
Therefore, $\Sigma_L$ can be dominant for most period of the impact
in the inviscid regime.

\bibitem{footnote3}Due to the discreteness of the measured $d$ values,
we use the calculated $d$ value at $t=t_{on}$ from the best fitting
function such as the solid line shown in Fig.3 lower inset.

\bibitem{footnote4} Actually $V_e$ also depends on $\mu$. But the dependence is
much weaker than $\sqrt\mu$ (unpublished data), and qualitatively
our argument is not affected. We also note that in Eq.~1, $V_e$
cancels out, thus its dependence on $\mu$ does not affect the
derivation of Eq.~2.

\end{thebibliography}
\end{document}